\documentclass[iop]{emulateapj}

\newcommand{\mytilde}{\raise.19ex\hbox{$\scriptstyle\sim$}}

\submitted{Accepted to ApJ (747:96)}

\shorttitle{HST Study of Dark Core in A520}
\shortauthors{Jee et al.}

\begin{document}

\title{A STUDY OF THE DARK CORE IN A520 WITH {\it HUBBLE SPACE TELESCOPE}:  THE MYSTERY DEEPENS\footnotemark[*]}

\footnotetext[*]{Based on observations made with the NASA/ESA {\it Hubble Space Telescope},
obtained at the Space Telescope Science Institute, which is operated by the Association of Universities for Research in Astronomy, Inc., 
under program 11221}

\author{M.~J.~JEE\altaffilmark{1}, A. MAHDAVI\altaffilmark{2}, H. HOEKSTRA\altaffilmark{3}, A. BABUL\altaffilmark{4},  J.~J.~DALCANTON\altaffilmark{5}, P. ~CARROLL\altaffilmark{5}, P.~ CAPAK\altaffilmark{6} }

\altaffiltext{1}{Department of Physics, University of California, Davis, One Shields Avenue, Davis, CA 95616, USA}
\altaffiltext{2}{Department of Physics and Astronomy, San Francisco State University, San Francisco, CA 94131, USA}
\altaffiltext{3}{Leiden Observatory, Leiden University, Leiden, The Netherlands}
\altaffiltext{4}{Department of Physics and Astronomy, University of Victoria, Victoria, BC, Canada}
\altaffiltext{5}{Department of Astronomy, University of Washington, Seattle, WA 98195, USA}
\altaffiltext{6}{Spitzer Science Center, California Institute of Technology, Pasadena, CA 91125, USA}

\begin{abstract}
We present a {\it Hubble Space Telescope}/Wide Field Planetary Camera 2  weak-lensing study of A520, where a previous analysis of ground-based data suggested the presence of a dark mass concentration. We map the complex mass structure in much greater detail leveraging
more than a factor of three increase in the number density of source galaxies available for lensing analysis.
The ``dark core'' that is coincident with the X-ray gas peak, but not with any stellar luminosity peak 
 is now detected with more than $10~\sigma$ significance.
The \mytilde1.5 Mpc filamentary structure elongated in the NE-SW direction is also clearly visible.
Taken at face value, the comparison among the centroids of dark matter, intracluster medium, and galaxy luminosity is at odds with what has been
observed in other merging clusters with a similar geometric configuration. To date, the most remarkable counter-example
might be the Bullet Cluster, which shows a distinct bow-shock feature as in A520, but no significant weak-lensing mass concentration around the
X-ray gas.
With the most up-to-date data, we consider several possible explanations that might lead
to the detection of this peculiar feature in A520.
However, we conclude that  none of these scenarios  can be singled out yet as the definite explanation for this puzzle.
\end{abstract}


\section{INTRODUCTION} \label{section_introduction}

Numerical simulations have successfully demonstrated that galaxy clusters form at the intersections of filaments and they grow by accreting other clusters/groups predominantly along the filaments. Because of the dominance of this virtually one-dimensional accretion, the cores of the galaxy clusters are subject to frequent near head-on collisions and thus are dynamically active. 

One of the key tools for studying merging clusters is the comparison among the distributions of the three cluster constituents: galaxies, hot plasma, and dark matter. For example, in merging clusters the intracluster medium suffers from ram pressure and lags behind galaxies and dark matter (e.g., Clowe et al. 2006; Jee et al. 2005a, 2005b), which are believed to be effectively collisionless. The contrast between collisional and collisionless components
becomes highest when we observe merging clusters at their core pass-through, when both the medium velocity and the effect of ram pressure stripping are largest.

The Bullet Cluster (Clowe et al. 2006) provides a remarkable example of the separation of cluster components. It possesses textbook examples of both bow shock and clear offsets between dark matter/galaxy and X-ray gas;  a numerical simulation of the cluster (Springel \& Farrar 2007) suggests that the subcluster (Bullet) is moving away from the main cluster at $\sim2700~\mbox{km}~\mbox{s}^{-1}$ with respect to the main cluster.
As the mass reconstruction is blind to the distribution of the cluster galaxies, the agreement of the mass clumps with the cluster galaxies and the offset of the X-ray gas from these are
strong evidence for collisionless dark matter (Clowe et al. 2006). 

Another merging cluster showing a comparably remarkable bow shock feature is A520 at $z=0.201$ (Markevitch et al. 2005). However, a weak-lensing study by Mahdavi et al. (2007, hereafter M07) finds a very perplexing mass structure. In addition to three mass clumps that are spatially coincident with the cluster galaxies, the mass reconstruction also
shows a significant ($\sim4\sigma$) mass peak that lies on top of the X-ray luminosity peak, but that is largely devoid of bright cluster galaxies.
Such a peculiar substructure, referred to as a ``Dark Core" in M07, does not appear in the weak-lensing mass map of the Bullet Cluster, which is believed to be at a similar merging stage.
M07 discussed several possible explanations such as background cluster contamination, bright galaxy ejection, line-of-sight (LOS) structure, or
violation of the upper limit. One extreme interpretation
would be that the dark matter particle's collisional cross section might be considerably larger than the upper limit ($\sigma_m/m_{DM} < 1~\mbox{cm}^{2}~\mbox{g}^{-1}$) derived from
the Bullet Cluster (Markevitch et al. 2004). Recently, independent results on the dark matter cross section supporting the collisionless nature 
have been reported from the studies of A2744 (Merten et al. 2011), MACS J0025.4--1222 (Bradac et al. 2008), DLSCL J0916.2+2951 (Dawson et al.  2011), halo ellipticities (Feng et al. 2010; see also Miralda-Escude 2002), etc. However, Williams \& Saha (2011) claim significant detection of light/mass offsets in A3827, which can be interpreted as evidence for collisional dark matter.

In this paper,  we present our weak-lensing study of A520 with {\it Hubble Space Telescope (HST) }/ Wide Field Planetary Camera 2 (WFPC2) images. 
This is a  critical follow-up study because high resolution imaging provides more usable galaxies for lensing analysis,
and thus enhances both the significance and resolution of the mass reconstruction.
These data verify the reality of the dark core and also refine the mass estimates of the substructures, which provide key ingredients for detailed numerical simulations.

We assume $(\Omega_M, \Omega_{\Lambda}, h) = (0.3, 0.7, 0.7)$ for cosmology unless explicitly stated otherwise. This gives a plate scale of $\mytilde3.3$~kpc/$\arcsec$ at the redshift ($z=0.201$)
of A520.
All the quoted uncertainties are at the 1$\sigma$ ($\mytilde68$\%) level.

\section{OBSERVATIONS AND DATA REDUCTIONS} \label{section_obs}

\begin{figure*}[th!]
\begin{center}   
\includegraphics[width=16cm]{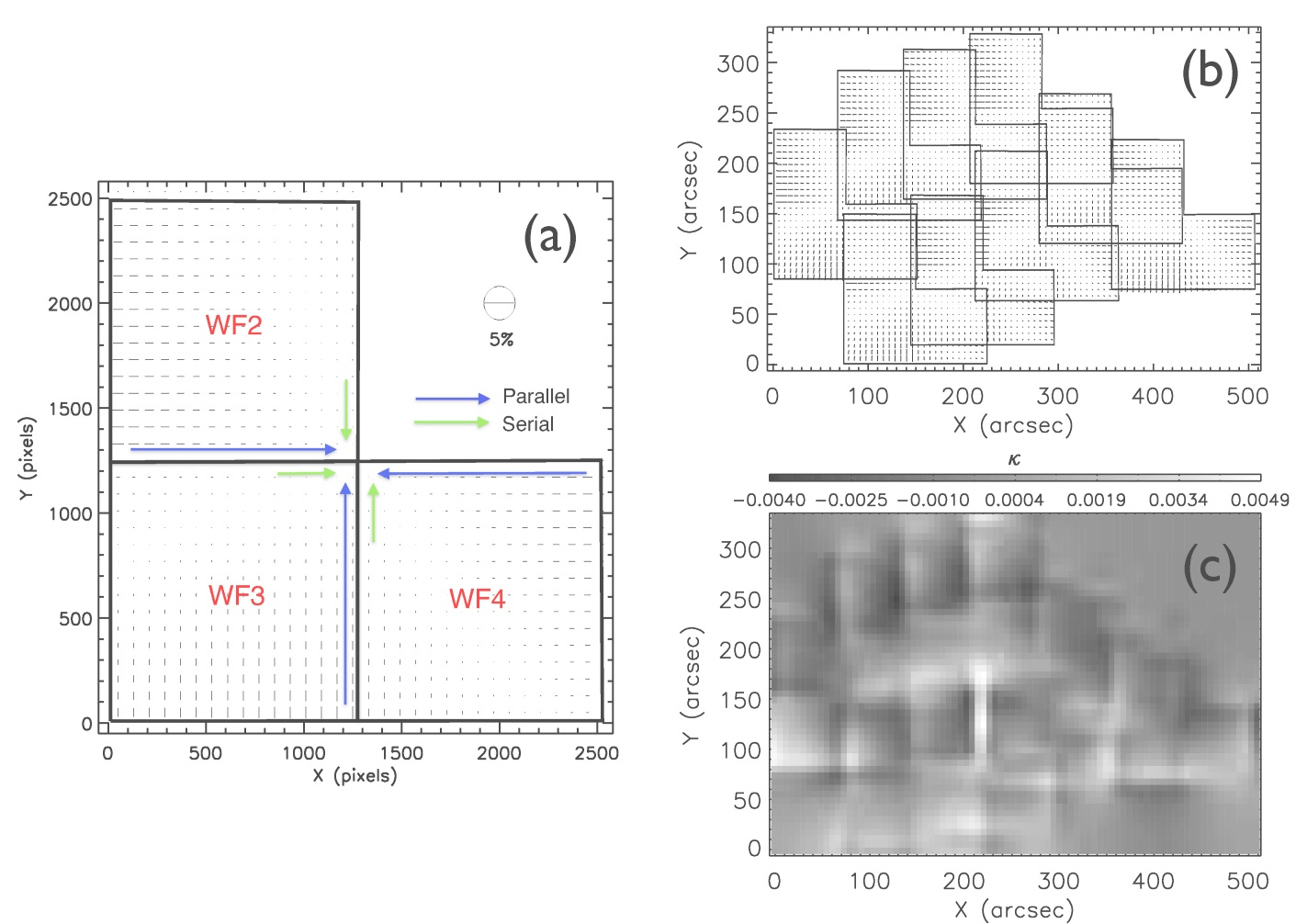}
\end{center}
\caption{Influence of CTI on mass reconstruction. (a) CTI pattern within WFPC2 CCDs.
It shows how typical circular WFPC2 PSFs are distorted due to CTI. The size and orientation of the ``whiskers'' represent the magnitude and direction of elongation, respectively.
(b) CTI-induced ellipticity pattern when the observational footprint is considered. (c) CTI-only mass reconstruction. 
The periodic variation due to the observational dither pattern is clear.  However, even the maximum amplitude of this uncorrected systematics for the average source galaxy is an order of magnitude smaller than the average lensing signal. 
\label{fig_wfpc2_cti_mass}}
\end{figure*}

The cluster A520 was observed in the F814W filter with WFPC2 in nine contiguous pointings covering approximately  the $8\farcm 5\times5\farcm5$
NE-SW elongated mass structure. Each pointing was dithered four times with a total integration of $4,400$ s. 
Our data reduction starts with the products of the STScI {\tt CALWP2} (Gonzaga et al. 2010) task. First, we ``drizzled'' (Fruchter \& Hook 2002)
each exposure individually to obtain rectified coordinates of objects to be used for alignment. Then, the alignment information
was fed into the {\tt MultiDrizzle} (Koekemoer et al. 2002) software to remove cosmic rays and create a distortion-free mosaic image. We chose a Gaussian interpolation kernel with a final
pixel size of $0\farcs06$ in order to reduce otherwise apparent aliasing due to the undersampled point-spread function (PSF). We discarded the data on the Planetary Camera (PC) for simplicity in data reduction,
and uniformity in angular resolution during the subsequent analysis.

The charge transfer inefficiency (CTI) of WFPC2 is potentially important because the deferred charge release smears and elongates shapes of objects, which must be
distinguished from the distortion by gravitational lensing. We measured the CTI by examining the ellipticity of sub-seeing features (cosmic rays, hot/warm pixels, etc.), 
which are non-astronomical objects and therefore not subject to the PSF of the instrument (Jee et al.  2011; hereafter J11).  We refer readers to J11 for technical
details. Figure~\ref{fig_wfpc2_cti_mass}(a) demonstrates how a hypothetical circular WFPC2 PSF placed at different positions would be distorted by CTI. Both parallel and serial CTI effects are found to be important while it is apparent that the effect in the parallel direction is much stronger. 

\begin{figure*}
\begin{center}
\includegraphics[width=15cm]{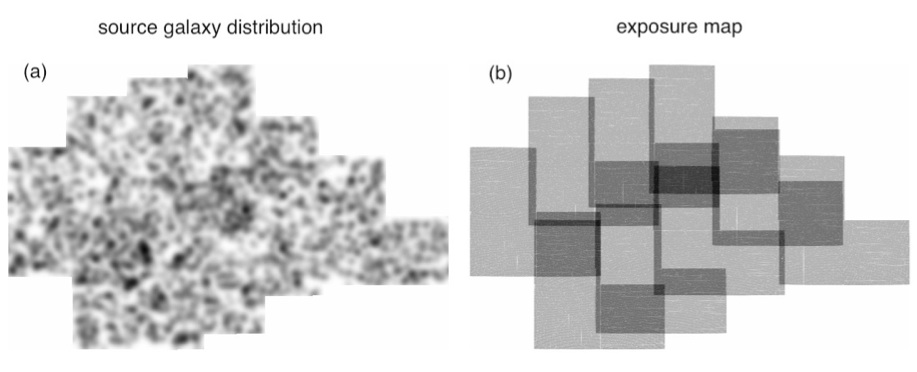}
\caption{Source galaxy distribution vs. exposure map. Darker shade represents higher value.
(a) Source galaxy positions are smoothed with an FWHM=10$\arcsec$ Gaussian kernel. (b) Shown here is the exposure map for the combined WFPC2 F814W image.  The comparison of the two figures hints at the correlation between the image depth and number density of source galaxies. However, the correlation is not strong.
On a $1\arcmin$ scale (the size of WF CCD), the peak-to-valley source galaxy number density variation is  at the $\sim7$\% level with respect to the mean density
whereas the exposure time varies significantly from $mytilde4,400$ s to $13,200$ s. Therefore, the spatial variation of the source galaxy number density 
due to the exposure time variation does not cause any spurious substructures in our mass reconstruction.
\label{fig_n_versus_exposure}}
\end{center}
\end{figure*}

The impact of this uncorrected CTI on our mass reconstruction is shown in Figures~\ref{fig_wfpc2_cti_mass}(b) and \ref{fig_wfpc2_cti_mass}(c).  Since no gravitational lensing signal is present here, the mass reconstruction would indicate potential impact of WFPC2 CTI on our mass reconstruction if no CTI correction were made. The periodic variation due to the observational dither pattern is clear.  However, even the maximum amplitude of this uncorrected systematics for the average source galaxy is an order of magnitude smaller than the average lensing signal. Therefore, the substructures in our mass reconstruction cannot be affected by any imperfect CTI correction.  

We construct a WFPC2 PSF library from archival images containing dense stellar fields (Jee et al. 2007a). Object ellipticity is determined by modeling the object surface profile with an elliptical Gaussian in the absence of the telescope seeing and the CTI. 
This is implemented by convolving the elliptical Gaussian with the model PSF (tweaked to simulate the CTI effect)  prior to fitting (J11).  We select our source galaxies whose F814W mag is fainter than 22 while removing spurious objects and cluster galaxies defined in M07 (see \textsection\ref{section_redshift} for details).  
The resulting number density is $\mytilde93$ galaxies $\mbox{arcmin}^{-2}$, approximately a factor of 3--4 higher than the number density of usable galaxies from the 
Canada-France-Hawaii-Telescope (CFHT) data. We display in Figure~\ref{fig_n_versus_exposure}(a) the distribution of these source galaxies. Comparison of the source distribution with the
exposure variation (Figure~\ref{fig_n_versus_exposure}(b)) hints at the correlation between the two maps. Nevertheless, this correlation is not to the extent that we worry about the
impact of the number density fluctuation on our mass reconstruction. On a $\sim1\arcmin$ scale (roughly the side of WF CCD), the peak-to-valley variation is at the $\sim7$\% level. 

\section{WEAK-LENSING ANALYSIS }\label{section_lensing}

\begin{figure*}
\begin{center}
\includegraphics[width=18cm]{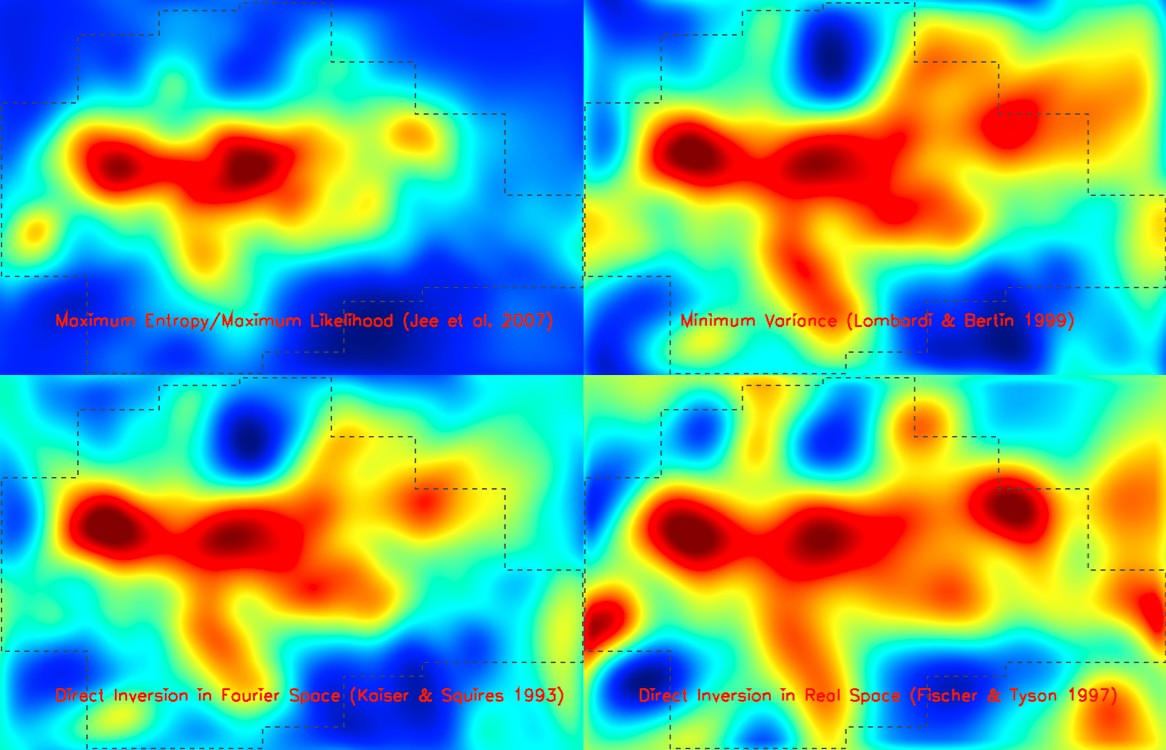}
\caption{Mass reconstruction in A520 with different algorithms. The dashed line shows the footprint of the WFPC2 observations. The structures well inside the boundary are all visible in different algorithms whereas the structures near and outside the boundary depend on the reconstruction algorithms.
\label{fig_mass_all}}
\end{center}
\end{figure*}

\begin{figure*}[t]
\begin{center}
\includegraphics[width=18cm]{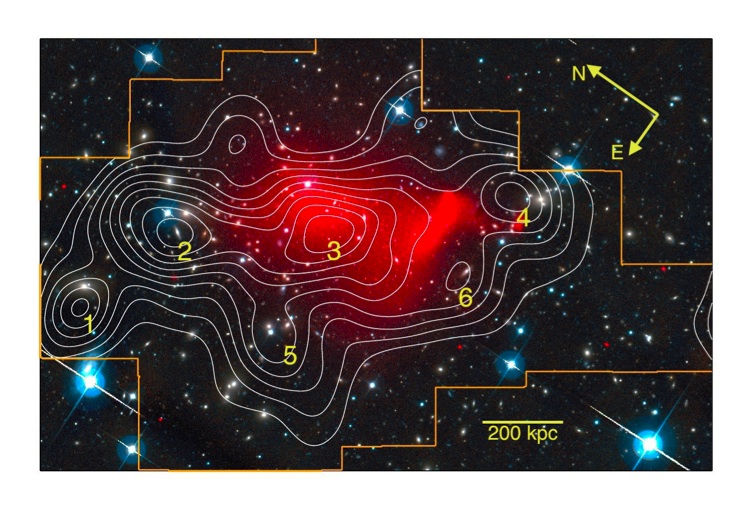}
\caption{Mass reconstruction in A520.  The intensity of the diffuse $Chandra$ emission is depicted in red. The background is the pseudo-color composite created from the CFHT $r$ and $g$ passband images.  The WFPC2 observation footprint is shown in orange. The white contours represent the convergence and the spacing is linear. The lowest contour corresponds to the 
$\sim2.6\sigma$ significance.
The numbers (1--6) indicate the significant mass peaks.
\label{fig_mass_chandra}}
\end{center}
\end{figure*}

\begin{figure*}
\begin{center}
\includegraphics[width=8cm]{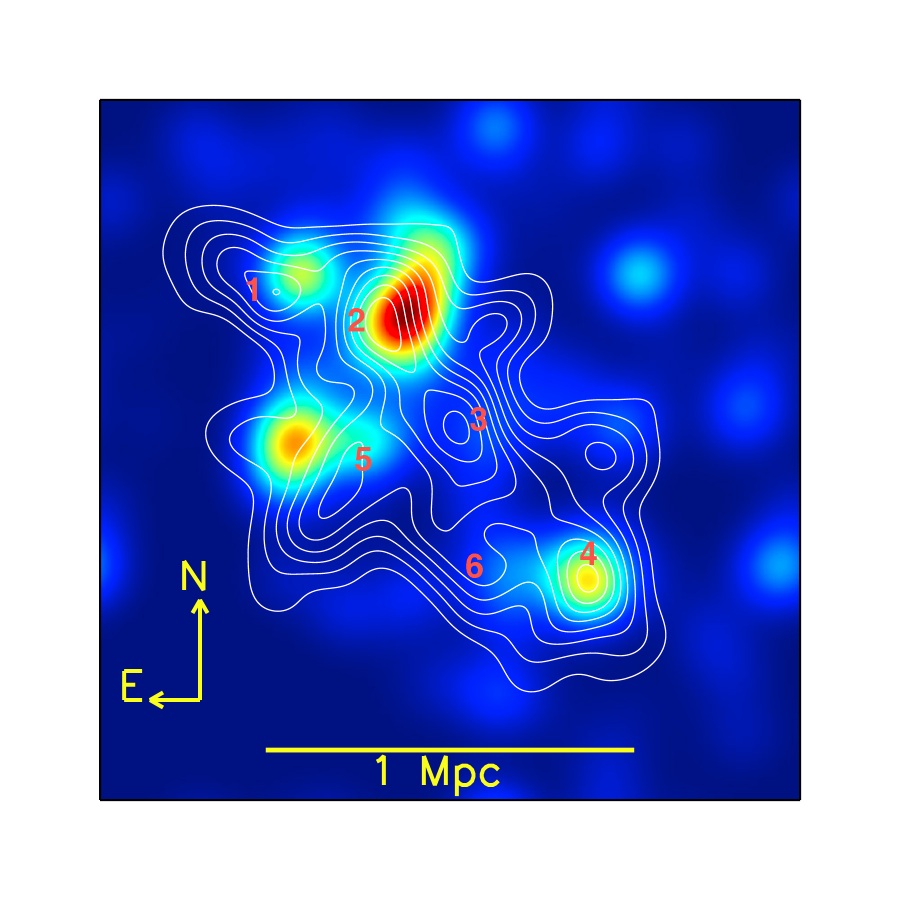}
\includegraphics[width=8cm]{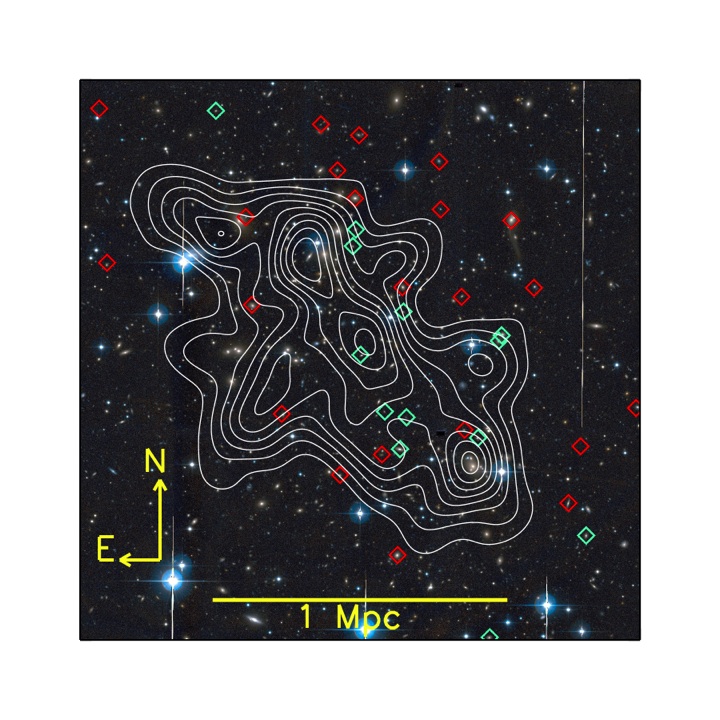}
\caption{Large-field mass reconstruction based on the combined ($HST$ and CFHT) catalogs. On the left-hand side, we overlay the mass contours on the smoothed rest-frame $B$-band luminosity distribution of the cluster (linear scale).
On the right-hand side, we illustrate the distribution of the high (red diamond, $\delta v_{rf}  > 1700~\mbox{km}~\mbox{s}^{-1}$) and low (green diamond, $\delta v_{rf} < -1500~\mbox{km}~\mbox{s}^{-1}$) velocity groups.
\label{fig_combined_mass}}
\end{center}
\end{figure*}

\subsection{Substructures}
We reconstruct the mass distribution using the maximum entropy/maximum likelihood (MEML) algorithm detailed in Jee et al. (2007b), which starts with a flat prior and then
updates the prior as the model improves.
We performed sanity checks with three other available codes (Kaiser \& Squires 1993;  Fischer \& Tyson 1997; Lombardi \& Bertin 1999) to examine if any pronounced substructure is also found in the other mass reconstruction algorithms. 
Figure~\ref{fig_mass_all} compares the results. The structures well inside the boundary (i.e., the footprint of the WFPC2 pointings) are all visible in different algorithms whereas the structures near and outside the boundary depend on the reconstruction algorithms. We observe that the MEML reconstruction minimizes the boundary effects most successfully. We also claim that the relative amplitudes for different mass peaks are most reliably recovered in this reconstruction because the algorithm does not pre-smooth the galaxy ellipticity and also it properly distinguishes shear ($\gamma$) from reduced shear 
[$\gamma/(1-\kappa)$] through iteration (important where the projected density is high). Therefore, hereafter our discussion on the substructure is based on the result from the MEML reconstruction.

We overlay the mass contours on the pseudo-color composite of A520 in Figure~\ref{fig_mass_chandra}. As in M07, we represent the intensity of the $Chandra$ X-ray emission
in red. The overall large-scale structure closely resembles the one reported in M07, although the current $HST$ data enable us to resolve the substructures with much higher significance. We label the six mass peaks following the scheme of M07 and refer to them as P1-6 hereafter.

The most remarkable consistency between the current and M07 mass maps is the presence of the strong ``dark core'' (P3), which coincides with the peak of the X-ray emission but lacks luminous cluster galaxies.
P1, P2, and P4 identified in M07 are in good spatial agreement with the equivalent peaks.  

P5 and P6 are new substructures identified in the current WFPC2 analysis. P5 was reported missing in M07, and this absence of a
distinct mass peak despite the apparent concentration of cluster galaxies was considered another unusual feature of M07 mass reconstruction.
The new mass map has therefore resolved this apparent discrepancy.
P6 coincides with some of the bright cluster galaxies, and the M07 mass map also indicated some overdensity in this region, albeit at a lower significance.

\begin{deluxetable*}{lcccccc}
\tabletypesize{\scriptsize}
\tablecaption{Mass and Luminosity Properties of Substructure ($r<150$~kpc)}
\tablenum{1}
\tablehead{\colhead{Substructure} & \colhead{$\alpha, \delta$} & \colhead{$\Delta \alpha, \Delta \delta$} & \colhead{Projected Mass} & \colhead{Luminosity} & \colhead{$M/L$} & \colhead{$f_g$}   \\
                    \colhead{}   & \colhead{ ($^h~^m~^s$, $\degr~ \arcmin~ \arcsec$)} & \colhead{ ($\arcsec,\arcsec$)}& \colhead{($h_{70}^{-1} 10^{13} M_{\sun}$)}  & \colhead{($h_{70}^{-2} 10^{11} L_{B \sun}$)}       & \colhead{($h_{70} M_{\sun}/L_{B \sun}$)}  & \colhead{$(h_{70}^{-1.5})$}  \\}
\tablewidth{0pt}
\startdata
P1  &  (04 54 20.76,  +02 57 38.4)   & (4.4,2.7) & $2.63\pm0.48$ &  $ 1.54 $  & $171\pm31 $ &  $  <0.06$\\
P2  & (04 54 15.02,  +02 57 09.2)    & (4.0,6.5) & $3.83\pm0.42$ &  $  3.58 $ & $106\pm12 $ &  $  <0.08 $\\
P3 (dark core) & (04 54 11.07, +02 55 35.3) & (6.7,6.5) & $4.00\pm0.38$ &  $ 0.68 $ & $ 588\pm56 $ &  $  <0.14$ \\
P4  & (04 54 04.32,  +02 53 51.0)  & (5.1,6.9)                   & $3.64\pm0.45$ &  $ 2.95$ & $123\pm15  $  &  $ <0.08$ \\
P5  & (04 54 16.53,  +02 55 26.7)  & (6.5,6.4)                   & $3.03\pm0.40$ &  $ 2.12$ & $143\pm19    $  &  $  <0.05$ \\
P6  & (04 54 08.85,  +02 53 50.2)  & (9.6,6.7)                   & $3.33\pm0.40$ &  $ 1.23$ & $270\pm33  $  &  $  <0.06 $\\
\enddata
\tablecomments{The positional uncertainty is estimated from bootstrapping.
We estimate the aperture mass based on the method of Fahlman et al. (1994). 
The mass uncertainties are evaluated from 1000 Monte-Carlo realizations.  
The gas fraction $f_g$ is derived using Cauchy-Schwartz method in M07
}
\end{deluxetable*}

\subsection{Mass, Luminosity, and Significance Estimation}
To estimate the cluster mass, we combined the shape catalogs from the current WFPC2 and the previous CFHT images by using WFPC2 shapes in the overlapping region and CFHT shapes for the rest. The resulting mass reconstruction is shown in Figure~\ref{fig_combined_mass}.
Since the WFPC2 images resolves fainter galaxies at higher redshift, the amplitude of the lensing signal in the $HST$ imaging is higher when no redshift correction is applied. Although the proximity of the cluster ($z=0.201$) makes the amount of correction small ($\lesssim6\%$), we take this into account in the mass estimation (see \textsection\ref{section_redshift} for details).
We find that the tangential shear profile at large radii ($r>200\arcsec$) is well described by a singular isothermal sphere (SIS) with 
$\sigma_v = 987\pm49~ \mbox{km}~\mbox{s}^{-1}$, which is consistent with the estimate of $1028\pm80~\mbox{km}~\mbox{s}^{-1}$ from M07. We estimate that the aperture mass (Fahlman et al. 1994) within $r=710$~kpc is
$(4.47\pm0.48)\times10^{14}~M_{\sun}$, again statistically consistent with the previous measurement $(5.00\pm0.55)\times10^{14}~M_{\sun}$.

For the determination of the $B$-band luminosity, we performed synthetic photometry using the Kinney et al. (1996) spectral templates, and derived a transformation to convert $g$-band (CFHT) magnitude and $g-r$ color to the rest frame $B$-band luminosity. The total luminosity within the $r=710$~kpc aperture is $1.7\times10^{12} L_{B \sun}$, which
gives a total mass-to-light ratio $263\pm28~M_{\sun}/L_{B \sun}$. The updated mass, $M/L$, and gas fraction for the substructures P1-6 are listed in Table 1.

The significance of the substructures is estimated with the following method. The background level cannot be determined reliably within the WFPC2 field (Figure~\ref{fig_mass_chandra}) where the cluster contribution is non-negligible. Thus, we created a wide-field convergence map from the WFPC2 + CFHT data and determined the
background level in the $13\arcmin< r < 17\arcmin$ annulus. The rms value of the convergence
at the location of the peak can be estimated in two ways. First, we can perform a bootstrapping analysis and compute the standard deviation with respect to the mean. Second, the Hessian
matrix (whose elements are the second derivatives of the likelihood function) can be utilized using the Gaussian approximation of the error distribution at the peak of the posterior distribution.
The first method is not a viable option in the current case because our high-resolution MEML mass reconstruction requires a significant CPU time ($\sim1$ day with 24 cores) to reach the
final solution. Therefore, in this paper we utilize the Hessian matrix to estimate the rms values of the mass pixels. Bridle et al. (1998) demonstrated that the errors derived from
their Hessian matrix are consistent with the values obtained from Monte Carlo simulations. We also confirmed their claim by performing independent
simulations using a low-resolution ($20\times 20$) mass grid.
We did not neglect the off-diagonal elements of 
the Hessian matrix because the mass pixels are highly correlated through the maximum-entropy regularization.

There is an ambiguity in determining the absolute convergence value at the location of the mass peak
because of the mass-sheet degeneracy. However, our wide-field mass reconstruction is 
performed in such a way that the convergence field in the $13\arcmin < r< 17\arcmin$ annulus is close 
to zero. Consequently, the convergence near the cluster mass peak remains virtually unchanged when we apply the $\kappa \rightarrow \lambda \kappa + (1-\lambda)$ transformation.

Within $r<150$ kpc, the significance (defined by the background-subtracted convergence divided by the rms)
of the ``dark core'' is $\sim12~\sigma$. This significance is slightly higher than
the value obtained from our aperture mass densitometry, which gives $\sim10~\sigma$. The difference in part comes from the fact that the aperture mass densitometry
uses a limited range of tangential shears. Nevertheless, we find that the ratio between the aperture mass and its uncertainty is a conservative measure of the significance of
the substructures (Table 1).

Another quantity related to the significance, but a different measure of the reliability of the mass peaks is their positional uncertainty. We performed a bootstrapping analysis by generating
1000 noise realizations and determine the centroid of each mass peak. Because our maximum-entropy method is slow and thus not practical for this experiment, we used
the Kaiser \& Squires (1993) algorithm.  The direction of the shifts during this bootstrapping run is nearly isotropic, and we show the $1\sigma$ value of the distribution in Table 1.
The typical $1\sigma$ deviation is $\sim8\farcs5$ corresponding to $\mytilde28$ kpc at the cluster redshift. 
Note that since the direct inversion by the Kaiser \& Squires (1993) is somewhat noisier than the MEML mass map, the estimated centroid
errors obtained in this way are likely to be overestimated (we also find that the significance of the dark core reduces to $\mytilde7\sigma$ when we use these 1000 noise realizations).
Nevertheless, these values are still small compared to the size of the mass peaks, which eliminates the possibility that the dark core
is the result of any catastrophic centroid shifts of nearby peaks.

\begin{figure}
\begin{center}
\includegraphics[width=8cm]{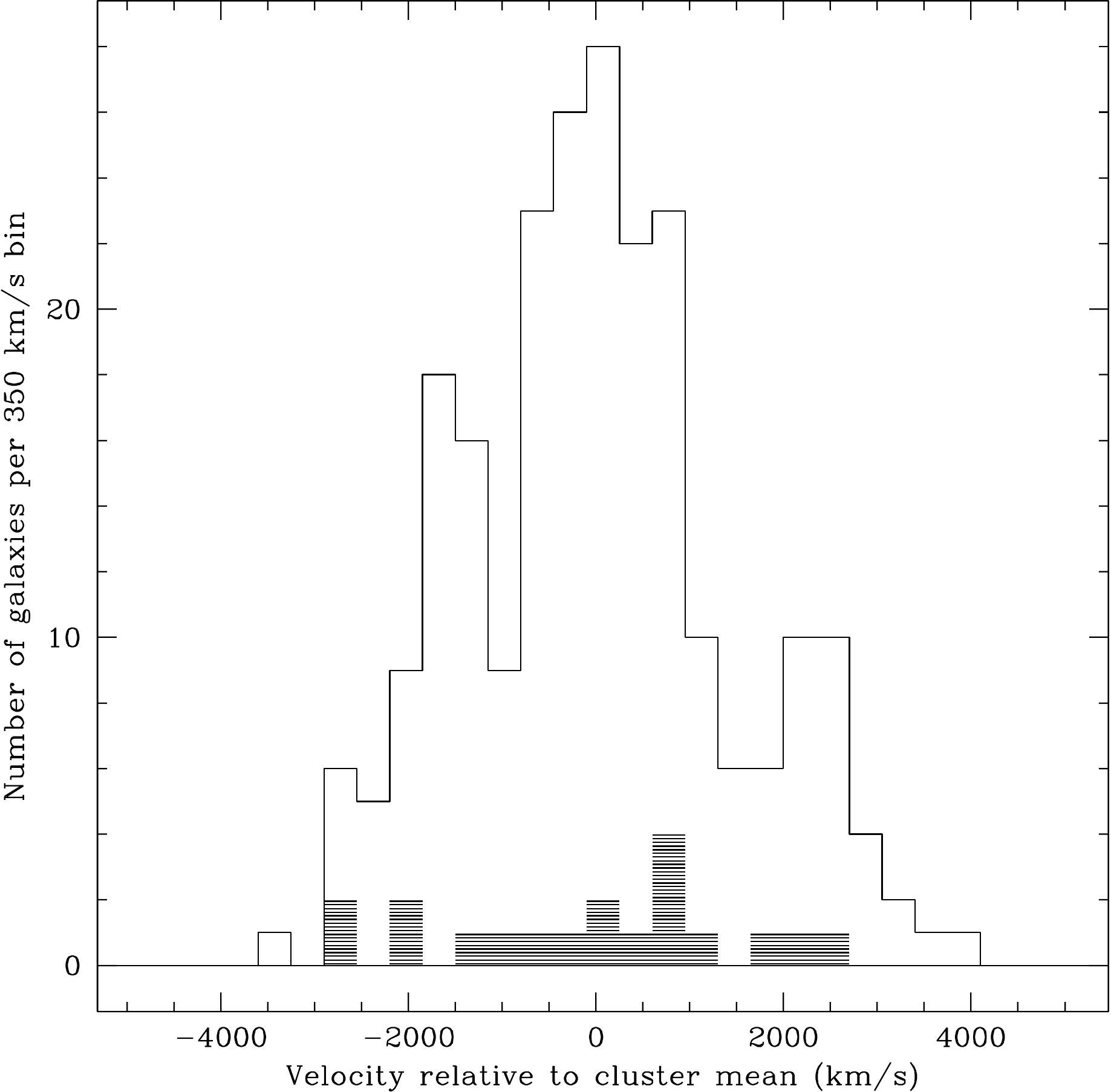}
\caption{Galaxy redshift distribution in the A520 field.
The distribution is inconsistent with a single Gaussian, and the redshift distribution shows a possible presence of high/low velocity groups.
However, we note that the spatial distribution of the galaxies belonging to the high/low velocity groups is not concentrated near the dark core, but instead scattered 
across the cluster field (see the right panel of Figure~\ref{fig_combined_mass}). The shaded histogram shows the redshift distribution of the galaxies
within the $r\sim150$ kpc radius of the dark core.
\label{fig_redshift_histo}}
\end{center}
\end{figure}

\subsection{Source Redshift Estimation} \label{section_redshift}

We select objects in the $22<F814W<27$ range as source galaxies.  
For the $22<F814W<24$ sources, we removed the red sequence galaxies 
defined by the CFHT data and also the spectroscopically confirmed cluster members.
The seeing of the CFHT A520 images is excellent, ranging from FWHM=$0\farcs50$ to $0\farcs57$. Hence, the
cross-identification between the WFPC2 and CHFT images can be done with high fidelity for this relatively bright subsample.
In the magnitude range $24<F814W<27$,  we did not attempt to remove cluster contamination.
At fainter magnitudes, most of the members will be blue, and thus the color-based cut becomes ineffective.
Also, in this regime the blue cluster galaxy contamination is low because the number of background galaxies overwhelms that of the cluster
galaxies, which is verified by our comparison of the number density in the WFPC2 field
with those from the Ultra Deep Field (UDF; Beckwith et al. 2003) and Great Observatories Origins Deep Survey (GOODS; Giavalisco et al. 2004) data.

The redshift distribution of our source galaxies is estimated using the publicly available UDF photo-$z$ catalog (Coe et al. 2006).
The method is explained in detail in our previous publications (e.g., Jee et al. 2005a; Jee \& Tyson 2009).
The critical lensing density $\Sigma_{crit}$ is proportional to
$\left < \beta \right >=\left < \mbox{max}(0, D_{ls}/D_s) \right >$, where $D_{ls}$ and $D_s$ are the angular
diameter distances between the lens and the source, and between the observer and the source, respectively.
For the galaxies available within the WFPC2 field,
we obtain $\left<\beta\right>=0.64$. Mahdavi et al. (2007) reported
that $\left<\beta\right>$ is $\sim0.60$ from their analysis with the Ilbert et al. (2006)
photometric redshift catalog. Therefore, when we combine the WFPC2 and CFHT shapes for the above mass estimation, we need to
correct for this $\mytilde6$\% difference in $\Sigma_{crit}$.

\section{NATURE OF THE DARK CORE}

Now that our high-resolution weak-lensing analysis confirms the dark core in A520, more extensive efforts should be made to understand the nature of this peculiar substructure.
In this paper, we review and extend the scenarios that may lead to this significant feature.

{\it Compact High M/L Group.}  Although we cannot find any giant elliptical galaxies near P3, there are at least $\sim11$ spectroscopic members within a $r=150$~kpc aperture. The smoothed light distribution (left panel of Figure~\ref{fig_combined_mass}) shows a marginal indication of a faint group coincident with the X-ray peak. However, if this is a group, the $M/L$ has to be extremely high, although the current $M/L$ estimate ($588\pm56~M_{\sun}/L_{B \sun}$) is somewhat lower than the previous value ($721\pm179~M_{\sun}/L_{B \sun}$).   One may suspect that the high concentration of the plasma near P3
somewhat contributes to this high $M/L$.  Because many lines of evidence (e.g., bow-shock feature) suggest that the concentration of the gas originated from other substructures (e.g., P2 and P4), it is worth examining the $M/L$ value for P3 in the absence of the gas.
Subtraction of the X-ray gas mass from the total mass reduces the $M/L$ value further to $\sim510~M_{\sun}/L_{B \sun}$
when the upper limit $0.52\times10^{13}~M_{\sun}$ of the plasma mass in M07 is assumed.
Nevertheless, this $M/L$ value is still substantially higher than the mean value of rich groups. For example, Parker et al. (2005) quote
$195\pm29~h_{70} M_{\sun}/L_{B \sun}$ for rich groups from their weak-lensing study of 116 groups.
However, considering that  
the distribution of the $M/L$ values of groups is still poorly known,
we believe that it is premature to rule out this possibility.

{\it Contribution from Neighboring Substructures.}  M07 investigated if a superposition of the two dark matter halos on P2 and P4 can lead to the
detection of P3. When we repeat the analysis with the current updated mass of the two halos and the addition of P5, the resulting aperture mass within a $r=150$~kpc would be
$\approx 3\times10^{12}~M_{\sun}$, still an order of magnitude smaller than what is required to produce the lensing signal.

{\it Distant Background Cluster.} An extremely massive cluster at significantly high redshift ($z>1$) can make its member galaxies difficult to be identified in the optical images, but can still signal its presence by distorting even higher redshift galaxies. However, the redshift dependence of the lensing signal is inconsistent with the object being at such high redshift. For example, when our mass reconstruction is performed with relatively bright source galaxies (thus mostly at $z<1$), the mass peak is still visible.

{\it Ejection of Bright Galaxies.}  With $N$-body simulations, Sales et al. (2007) find that a significant fraction of satellite galaxies are ejected during their first approach to the main system via three-body encounter.  M07 considered the possibility that the dark core in A520 became devoid of bright cluster members by the same dynamical process. This hypothesis is in part motivated by the presence of the rich galaxy group  (P5) $\sim300$~kpc east of the dark core. There was no significant mass associated with P5 in M07. However, as mentioned in \textsection\ref{section_lensing}, our new analysis reveals the clear peak on top of the galaxy group. Moreover, the ejected component in the three-body encounter is always the weaker system of the pair infalling to the main system. Therefore, we rule out this possibility for the peculiar structure of A520.

{\it Collisional Deposition of Dark Matter.}  The X-ray analysis of the prominent bow shock feature (Markevitch et al. 2005)
indicates that perhaps the subcluster (P4) was passing through the main cluster's core at a supersonic speed.
If we attribute P3 to a deposition of collision-stripped dark matter, indeed A520 is a counterexample to the Bullet Cluster. 
Assuming that P1, P2, P4, and P5 equally contribute to the total mass of P3, M07 estimates
$\sigma_{DM}/m_{DM} \approx 3.8\pm1.1~ \mbox{cm}^2\mbox{g}^{-1}$, which exceeds the upper limit 
$\sigma_{DM}/m_{DM} <1 ~ \mbox{cm}^2~\mbox{g}^{-1}$ derived from the Bullet Cluster (Markevitch et al. 2004), although we note that Williams \& Saha (2011) suggest that a kpc-scale separation between stellar and dark matter components in the cluster A3827 may be evidence for dark matter with a non-negligible self-interaction cross-section.
Our current WFPC2-based study reduces the mass
uncertainty of the substructures by more than a factor of 2 while the mass estimates remain close to the previous measurements. Therefore, the current improvement in precision 
only increases the significance of the above discrepancy.

{\it Filament along the LOS direction} Numerical simulations show that galaxy clusters form where filaments intersect. Of course, the most apparent filament in A520 
might be the NE--SW structure traced by galaxies, X-rays, and mass. If another filament associated with A520 is oriented along the LOS, its center should perhaps be
near P3, because it corresponds to the approximate center of the large scale structure even with P3 excluded.  As M07 noted, this filament should be sparse enough that no strong concentration of X-ray gas nor galaxies is observable. From the velocity field analysis with 167 spectroscopic cluster members, Girardi et al. (2008) report possible existence of high velocity group, whose rest-frame LOS velocity is $\sim2000~\mbox{km}~\mbox{s}^{-1}$ relative to the main system, which can be viewed as indicating a LOS filament.

Our analysis with 236 redshifts obtained from the Canadian Network for Observational Cosmology (CNOC1; Yee et al. 1996) redshift survey and 
our independent DEIMOS redshift survey of the cluster field also supports the possibility of the aforementioned high velocity group along the LOS direction.
Figure~\ref{fig_redshift_histo} shows the redshift distribution in the A520 field. The presence of the distinct bump at $v_{rf}\sim2000~\mbox{km}~\mbox{s}^{-1}$ is clear. Moreover, the low end of the distribution possesses marginal  indication of a possible presence of a low velocity group 
at $v_{rf}\sim-1500~\mbox{km}~\mbox{s}^{-1}$.  

However, the spatial distributions of both high and low velocity groups are not compact, but rather scattered, roughly following the large scale structure (see the right panel of Figure~\ref{fig_combined_mass}) , which poses a challenge to the scenario that these groups are responsible for the detection of the dark core.
If both groups are dynamically associated with the main cluster, the high and low velocity groups might be loose foreground and background groups, respectively, in-falling to the main system along the LOS direction. The velocity dispersion of the high velocity group is $418\pm63~\mbox{km}~\mbox{s}^{-1}$ estimated out of 34 galaxies whose rest-frame velocities are $>1500~\mbox{km}~\mbox{s}^{-1}$ from the center of the main system. If the group's mass is assumed to follow
a SIS, the projected mass within an $r=150$~kpc aperture is $\sim2\times10^{13}~M_{\sun}$. Therefore, if we further assume that the low velocity group contains a similar mass, the contribution
from the two LOS groups and the X-ray gas ($\lesssim0.5\times10^{13}~M_{\sun}$) can marginally add up to the required mass of the dark core.
As a matter of course, one questionable assumption is that the cross-sectional mass profile of the filament has a steep gradient, the plausibility of which requires detailed numerical studies.
The galaxy distributions of the high and low velocity groups do not show any indication of such a gradient.
If the hypothesized LOS filament possesses a smooth cylindrical mass distribution (i.e., without central cuspiness), it is not likely to lead to the
detection of any isolated substructure as the dark core.
Another immediate problem is that the filament should have increased the $M/L$ values of other substructures as well, inconsistent with our measurements (Table 1).

Aside from the above diffuse filament scenario, one can also imagine a thin filament coincident with P3. In this case we do not associate the aforementioned high velocity group with this narrow filament. Instead, we suggest that the 11 cluster galaxies mentioned in the context of the compact high $M/L$ group represent group-like substructure embedded in this narrow filament.  This mass configuration can explain the lensing detection of the dark core, but it too poses concerns. First, while the additional mass in the filament would result in a high $M/L$ for the substructure, an $M/L$ of $\sim510~M_{\sun}/L_{B \sun}$ is considerably larger than the typical observed $M/L$ of $\sim300~M_{\sun}/L_{B \sun}$ for filaments (Eisenstei et al. 1997; Schirmer et al. 2011). Second, the hypothesis is hard to prove (or disprove) given the current statistics derived from $\sim11$ galaxies.

\section{SUMMARY AND CONCLUSIONS}  \label{section_conclusion}

We have presented our $HST$/WFPC2 lensing study and confirmed the reality of the peculiar mass structure of A520.
More than a factor of three increase in the number of usable galaxies per unit area enables us to trace the complex mass distribution of the cluster with higher resolution and with greater significance. The ``dark core'' that is coincident with the location of the X-ray luminosity peak but is largely devoid of bright cluster galaxies is
clearly detected at the $>10\sigma$ level. With the current data,
we reviewed several scenarios which may explain the presence of this unusual mass distribution.

Dark matter self-interaction cross-section must be  at least $\mytilde6~\sigma$ larger than the upper limit $\mytilde1\mbox{cm}^2\mbox{g}^{-1}$ determined by the Bullet Cluster
observation. Therefore, it is difficult to attribute the feature to dark matter self-interaction without falsifying the weak lensing analysis of the Bullet Cluster (Clowe et al. 2006), which
does not show any significant mass clump between the two dominant mass peaks.
Interestingly, when Bradac et al. (2006) combined the strong- and weak-lensing signals, they revealed a non-negligible mass concentration coincident with the main X-ray peak.
Therefore, it might be worth investigating further the reality of this feature of the Bullet Cluster in detail.
We also note that the estimate $\mytilde3.8\pm1.1~\mbox{cm}^2\mbox{g}^{-1}$ by M07 is still within the upper limits set by other dark matter collisional cross-section studies
(e.g., Natarajan et al. 2002; Hennawi \& Ostriker 2002; Bradac et al. 2008; Merten et al. 2011).

The presence of a galaxy group with unusually high $M/L$ might be a plausible solution, although
the required $M/L$ is  considerably higher than the mean value for very rich groups. Since the distribution of the $M/L$ values of galaxy groups is still poorly known,
we cannot exclude this possibility yet.

We also considered a scenario, wherein a fortuitous superposition of an LOS filament is located near the dark core.
This hypothesis is in part supported by the spectroscopic data. If the velocity dispersion of the high velocity group is representative of its mass and if the low velocity group contains a similar mass, it is
possible that the sum of the two masses  can cause the lensing detection of the dark core. However, the spatial distributions of both groups are very broad and not concentrated on the dark core.
Hence, it is difficult to rationalize that the projected density profile of the hypothesized diffuse LOS filament possesses such a cuspiness at its center. 

Alternatively, we also discussed possible existence of a thin filament coincident with the dark core. Although this might explain the lensing detection of such a peculiar substructure, the current data (only 11 spectroscopic redshifts) do not provide sufficient statistics to convince us of the presence of such a thin, but long filament near the dark core.
More extensive spectroscopic surveys near the dark core area are required
to test the hypothesis. 

Despite our solid confirmation on the presence of the dark core, we conclude that it is yet premature to single out the most probable cause of the dark core from the above
scenarios.
\\

H. Hoekstra acknowledges support from the Netherlands Organisation for ScientiÞc Research (NWO) through a VIDI grant. 
H. Hoekstra is also supported by a Marie Curie International Reintegration Grant. A. Babul acknowledges support from an NSERC Discovery grant.
J. Dalcanton and A. Mahdavi acknowledge the support by NASA through program GO-11221.


\begin{thebibliography}{}
\bibitem[Brada{\v c} et al.(2006)]{2006ApJ...652..937B} Brada{\v c}, M., 
Clowe, D., Gonzalez, A.~H., et al.\ 2006, \apj, 652, 937 
\bibitem[Brada{\v c} et al.(2008)]{2008ApJ...687..959B} Brada{\v c}, M., 
Allen, S.~W., Treu, T., et al.\ 2008, \apj, 687, 959 
\bibitem[Beckwith, Somerville, \& Stiavelli(2003)]{beckwith03} Beckwith, S., Somerville, R., Stiavelli M., 2003, STScI Newsletter vol 20 issue 04
\bibitem[Bridle et al.(1998)]{1998MNRAS.299..895B} Bridle, S.~L., Hobson, 
M.~P., Lasenby, A.~N., \& Saunders, R.\ 1998, \mnras, 299, 895 
\bibitem[Clowe et al.(2006)]{2006ApJ...648L.109C} Clowe, D., Brada{\v c}, 
M., Gonzalez, A.~H., Markevitch, M., Randall, S.~W., Jones, C., 
\& Zaritsky, D.\ 2006, \apjl, 648, L109 
\bibitem[Coe et al.(2006)]{2006AJ....132..926C} Coe, D., Ben{\'{\i}}tez,
N., S{\'a}nchez, S.~F., Jee, M., Bouwens, R., \& Ford, H.\ 2006, \aj, 132,
926
\bibitem[Dawson et al.(2011)]{2011arXiv1110.4391D} Dawson, W.~A., Wittman, 
D., Jee, M., et al.\ 2011, arXiv:1110.4391 
\bibitem[Fahlman et al.(1994)]{fahlman94} Fahlman,G.,Kaiser,N.,Squires,G.\& Woods,D. 1994,\apj, 437, 56
\bibitem[Fischer 
\& Tyson(1997)]{1997AJ....114...14F} Fischer, P., \& Tyson, J.~A.\ 1997, \aj, 114, 14 
\bibitem[Fruchter \& Hook(2002)]{2002PASP..114..144F} Fruchter, A.~S., \&
Hook, R.~N.\ 2002, \pasp, 114, 144
\bibitem[Giavalisco et al.(2004)]{giavalisco04} Giavalisco, M., et al.\ 2004, \apjl, 600, L93
\bibitem[Girardi et 
al.(2008)]{2008A&A...491..379G} Girardi, M., Barrena, R., Boschin, W., \& Ellingson, E.\ 2008, \aap, 491, 379 
\bibitem[Gonzaga et al. (2010)]{2010wfpc2handbook} Gonzaga, S., \& Biretta, J., et al. 2010, in HST WFPC2 Data Handbook, v. 5.0,ed., Baltimore, STScI
\bibitem[Eisenstein et al.(1997)]{1997ApJ...475..421E} Eisenstein, D.~J., 
Loeb, A., \& Turner, E.~L.\ 1997, \apj, 475, 421 
\bibitem[Feng et al.(2010)]{2010PhRvL.104o1301F} Feng, J.~L., Kaplinghat, 
M., \& Yu, H.-B.\ 2010, Physical Review Letters, 104, 151301 
\bibitem[Ilbert et
al.(2006)]{2006A&A...457..841I} Ilbert, O., et al.\ 2006, \aap, 457, 841
\bibitem[Jee et al.(2005a)]{2005ApJ...618...46J} Jee, M.~J., White, R.~L.,
Ben{\'{\i}}tez, N., Ford, H.~C., Blakeslee, J.~P., Rosati, P., Demarco, R.,
\& Illingworth, G.~D.\ 2005a, \apj, 618, 46
\bibitem[Jee et al.(2005b)]{2005ApJ...634..813J} Jee, M.~J., White, R.~L.,
Ford, H.~C., Blakeslee, J.~P., Illingworth, G.~D., Coe, D.~A., \& Tran,
K.-V.~H.\ 2005b, \apj, 634, 813
\bibitem[Jee et al.(2007)]{2007PASP..119.1403J} Jee, M.~J., Blakeslee, 
J.~P., Sirianni, M., Martel, A.~R., White, R.~L., 
\& Ford, H.~C.\ 2007a, \pasp, 119, 1403
\bibitem[Jee et al.(2007)]{2007ApJ...661..728J} Jee, M.~J., et al.\ 2007b, 
\apj, 661, 728 
\bibitem[Jee \& Tyson(2009)]{2009ApJ...691.1337J} Jee, M.~J., \& Tyson, J.~A.\ 2009, \apj, 691, 1337
 \bibitem[Jee et al.(2011)]{2011ApJ...737...59J} Jee, M.~J., et al.\ 2011, 
\apj, 737, 59 
\bibitem[Hennawi 
\& Ostriker(2002)]{2002ApJ...572...41H} Hennawi, J.~F., \& Ostriker, J.~P.\ 2002, \apj, 572, 41
\bibitem[Kaiser \& Squires(1993)]{ks93} Kaiser, N.~\& Squires, G.\ 1993, \apj, 404, 441 
\bibitem[Kinney et al.(1996)]{1996ApJ...467...38K} Kinney, A.~L., Calzetti, 
D., Bohlin, R.~C., McQuade, K., Storchi-Bergmann, T., 
\& Schmitt, H.~R.\ 1996, \apj, 467, 38 
\bibitem[Koekemoer et al.(2002)]{2002multidrizzle} Koekemoer, A. M., Fruchter, A. S., Hook, R. N., \& Hack, W. 2002, in The 2002
HST Calibration Workshop, ed. S. Arribas, A. Koekemoer, \& B. Whitmore (Baltimore: STScI), 337
\bibitem[Lombardi \& Bertin(1999)]{lb99} Lombardi, M.~\& Bertin, G.\ 1999, \aap, 348, 38
\bibitem[Mahdavi et al.(2007)]{2007ApJ...668..806M} Mahdavi, A., Hoekstra, 
H., Babul, A., Balam, D.~D., \& Capak, P.~L.\ 2007, \apj, 668, 806 
\bibitem[Markevitch et al.(2002)]{2002ApJ...567L..27M} Markevitch, M., 
Gonzalez, A.~H., David, L., Vikhlinin, A., Murray, S., Forman, W., Jones, 
C., \& Tucker, W.\ 2002, \apjl, 567, L27 
\bibitem[Markevitch et al.(2004)]{2004ApJ...606..819M} Markevitch, M., 
Gonzalez, A.~H., Clowe, D., Vikhlinin, A., Forman, W., Jones, C., Murray, 
S., \& Tucker, W.\ 2004, \apj, 606, 819 
\bibitem[Markevitch et al.(2005)]{2005ApJ...627..733M} Markevitch, M., 
Govoni, F., Brunetti, G., \& Jerius, D.\ 2005, \apj, 627, 733 
\bibitem[Merten et al.(2011)]{2011MNRAS.417..333M} Merten, J., Coe, D., 
Dupke, R., et al.\ 2011, \mnras, 417, 333 
\bibitem[Miralda-Escud{\'e}(2002)]{2002ApJ...564...60M} Miralda-Escud{\'e}, 
J.\ 2002, \apj, 564, 60 
\bibitem[Natarajan et al.(2002)]{2002ApJ...580L..17N} Natarajan, P., Loeb, 
A., Kneib, J.-P., \& Smail, I.\ 2002, \apjl, 580, L17 
\bibitem[Parker et al.(2005)]{2005ApJ...634..806P} Parker, L.~C., Hudson, 
M.~J., Carlberg, R.~G., \& Hoekstra, H.\ 2005, \apj, 634, 806 
\bibitem[Sales et al.(2007)]{2007MNRAS.379.1475S} Sales, L.~V., Navarro, 
J.~F., Abadi, M.~G., \& Steinmetz, M.\ 2007, \mnras, 379, 1475 
\bibitem[Schirmer et 
al.(2011)]{2011A&A...532A..57S} Schirmer, M., Hildebrandt, H., Kuijken, K., \& Erben, T.\ 2011, \aap, 532, A57 
\bibitem[Springel 
\& Farrar(2007)]{2007MNRAS.380..911S} Springel, V., \& Farrar, G.~R.\ 2007, \mnras, 380, 911 
\bibitem[Williams 
\& Saha(2011)]{2011MNRAS.415..448W} Williams, L.~L.~R., \& Saha, P.\ 2011, \mnras, 415, 448 
\bibitem[Yee et al.(1996)]{1996ApJS..102..269Y} Yee, H.~K.~C., Ellingson, 
E., \& Carlberg, R.~G.\ 1996, \apjs, 102, 269 
\end{thebibliography}
\end{document}